**Research Article**

# Reducing Cybersickness in 360-degree Virtual Reality

Iqra Arshad[1], Paulo De Mello[2], Martin Ender[2], Jason D. McEwen[2], Elisa R. Ferré[1,3]

1 Department of Psychology, Royal Holloway University of London, Egham, UK

2 Kagenova Limited, Guildford GU5 9LD, UK

3 Department of Psychological Sciences, Birkbeck University of London, London, UK

**Corresponding Author:**

Elisa R. Ferré

Department of Psychology

Royal Holloway University of London

Egham, Surrey TW20 0EX, UK

e-mail: e.ferre@rhul.ac.uk

tel: +44 (0) 1784 443530

**Abstract**

Despite the technological advancements in Virtual Reality (VR), users are constantly combating feelings of nausea and disorientation, the so called *cybersickness*. Cybersickness symptoms cause severe discomfort and hinder the immersive VR experience. Here we investigated cybersickness in 360-degree head-mounted display VR. In traditional 360-degree VR experiences, translational movement in the real world is not reflected in the virtual world, and therefore self-motion information is not corroborated by matching visual and vestibular cues, which may trigger symptoms of cybersickness. We have evaluated whether a new Artificial Intelligence (AI) software designed to supplement the 360-degree VR experience with artificial 6-degrees-of-freedom motion may reduce cybersickness. Explicit (simulator sickness questionnaire and fast motion sickness rating) and implicit (heart rate) measurements were used to evaluate cybersickness symptoms during and after 360-degree VR exposure. Simulator sickness scores showed a significant reduction in feelings of nausea during the AI supplemented 6-degrees-of-freedom motion VR compared to traditional 360-degree VR. However, 6-degrees-of-freedom motion VR did not reduce oculomotor or disorientation measures of sickness. No changes have been observed in fast motion sickness and heart rate measures. Improving the congruency between visual and vestibular cues in 360-degree VR, as provided by the AI supplemented 6-degrees-of-freedom motion system considered, is essential to provide a more engaging, immersive and safe VR, which is critical for educational, cultural and entertainment applications.

# 1. Introduction

As we move through the external environment, the vestibular system in the inner ear provides a flow of information regarding the position of our head in the three-dimensional space. The vestibular system comprises of three orthogonal semicircular canals (anterior, posterior and horizontal), that respectively sense rotational acceleration of the head in space and around the cardinal yaw, roll, and pitch axes, and two otolith organs (utricle and saccule) that sense translational acceleration, including the orientation of the head relative to gravity. These signals are crucial for self-motion perception (Chen *et al.,* 2008) confirming the essential vestibular contribution to the brain's *Global Positioning System*.

Typically, vestibular inputs are perfectly matching with optic flow information from the images moving across the retina. The brain optimally integrates vestibular and visual signals to create a coherent perception of the direction and speed of self-motion (Butler *et al*., 2010; Fetsch *et al*., 2009; Greenlee *et al*., 2016). However, there are some circumstances in which visual and vestibular signals for self-motion may not match and potentially conflict. This is the case of Virtual Reality (VR). The promise of VR has always been enormous and the ability to immerse a user in a virtual environment through the use of 3D real-time computer graphics and advanced head-mounted display devices has been shown to be beneficial in a number of applications like engineering, education and entertainment. VR has grown exponentially in popularity in recent years. Improvements in resolution, latency, flicker rate and motion tracking have enhanced the VR experience. However, a troubling problem remains where up to 80% of VR users experience debilitating symptoms of discomfort, disorientation, nausea, eyestrain, headaches and sweating, a malady called *cybersickness* (LaViola, 2000; Rebenitsch and Owen, 2016). Cybersickness is characterised by severe and frequent disorientation symptoms (dizziness, vertigo, and difficulty in focusing), followed by nausea symptoms (stomach awareness, increased salivation, and nausea itself), and least oculomotor symptoms (eyestrain, headache, and blurred vision), the so-called D>N>O profile (Rebenitsch and Owen, 2016; Stanney and Kennedy, 1997). This symptom profile distinguishes cybersickness from other types of motion sickness, such as simulator sickness and sea sickness. Critically, cybersickness may hinder the immersive and interactive qualities of VR and its real-life applications (Yildirim, 2019).

Although its causes are not entirely clear, cybersickness may be due to a discrepancy between vestibular and visual information about body's orientation and self-motion (Gallagher and Ferrè, 2018; Reason and Brand, 1975; Reason, 1978). In typical VR scenarios, such as a VR driving simulator, the simulation provides accurate optic flow patterns of the road, surrounding buildings, people and other parts of the environment, which elicit visual information of self-motion (*vection*; So *et al.*, 2001). The visual signals tell the brain that the user is moving with a certain speed and in a certain direction. However, since the user is not actually moving, the vestibular organs signal zero angular or translational acceleration. Thus, if the perception of vection is not corroborated by self-motion signals transmitted by the vestibular system, a sensory conflict is likely to occur, and cybersickness may ensue. Accordingly, higher levels of cybersickness have been described in flight simulators and in VR games in which a high level of sensory conflict is present. However, other accounts of cybersickness have been suggested. For instance, it has been proposed that the provocative stimulus is the postural instability triggered in VR (Stoffregen, 1991; Stoffregen, 2014). Additionally, the sensory conflict account has been refined by some to focus mainly on mismatches between the estimated and true environmental vestibular vertical (Bos *et al*., 2008). A common feature of these accounts seems to be that perceptual systems are highly reliable and accurate and VR technology challenges the senses and their integration.

Recent studies have shown promising results in reducing cybersickness by visuo-vestibular sensory re-coupling methods. Cybersickness improves when the absent vestibular information during simulated self-motion are mimicked by applying natural or artificial vestibular stimulation. For instance, using a rotational chair and timely coupling of physical movement with VR effectively reduced cybersickness symptoms (Ng *et al*., 2020). Similarly, cybersickness has been shown to improve with Galvanic Vestibular Stimulation (GVS) to artificially mimic vestibular cues in VR (Gallagher *et al*., 2020; Weech *et al*., 2018).

However, research has mainly focused on visuo-vestibular conflicts triggered by VR environments in which vestibular signals are absent. That is, when VR users feel the sensation of travelling through a virtual environment, while actually remaining stationary in the real world. Little is known on visuo-vestibular conflicts in VR scenarios in which the user is allowed to move, such as in *360-degree VR*. 360-degree VR refers to videos or photos that are captured using specialist omnidirectional cameras comprised of multiple lens that enable filming of a full panoramic view of 360 degrees. The user is therefore able to look around the entire scene and perceive an immersive experience. However, in almost all 360-degree VR experiences, movement in the real physical world is not reflected in the virtual world. In other words, current-generation 360-degree VR supports only 3-degrees-of-freedom rotational motion where the virtual scene reflects changes in the rotation of the user's head, but does not support translational motion. Thus, VR lacks the 6-degrees-of-freedom self-motion that we normally experience in the real world provided by multisensory integration between dynamic visual and 6-degrees-of-freedom vestibular cues, as well as the within channel integration between angular acceleration cues sensed by the vestibular semicircular canals and translational acceleration sensed by the otolith organs. Perhaps not surprisingly, the sensory mismatch in traditional 360-degree VR may easily induce cybersickness symptoms. Many studies have reported various physical symptoms such as headaches, focusing difficulties and dizziness during 360-degree VR viewing (Carnegie and Rhee, 2015; Elwardy *et al.*, 2020; Kim *et al.*, 2019; Kim *et al.*, 2019; Padmanaban *et al.*, 2018).

Here, we consider cybersickness in 360-degree head-mounted display VR. We systematically explored the cybersickness profile in settings where there are minimal dynamic visual inputs combined with translational vestibular inputs. We also investigate the effectiveness of a new Artificial Intelligence (AI) based software in reducing cybersickness. Where previous VR applications have focused on a 3-degrees-of-freedom rotational motion, this software aims to reduce cybersickness through providing both rotational and translational motion. We predict that this AI based software may improve the congruency between visual and vestibular signals during VR exposure and enhance realism. This may have a potential impact on immersive VR experience with fewer unwanted side effects of cybersickness.

## 2. Methods

### 2.1. Ethics

The experimental protocol was approved by the ethics committee at Royal Holloway, University of London. The experiment was conducted in line with the Declaration of Helsinki. Written informed consent was obtained prior to commencing the experiment.

### 2.2. Participants

Twenty-five healthy right-handed participants volunteered in the study (19 women; mean age= 22.16 years, SD = 7.12 years). The sample size was estimated a priori based on similar experimental procedures (Gallagher *et al.*, 2019; Gallagher *et al.*, 2020), set in advance of testing and was also used as data-collection stopping rule. Participants with a history of neurological, psychiatric, vestibular or auditory disorders were excluded.

### 2.3. An AI-based software to provide 6-degree-of-freedom Virtual Reality experience

An AI-based software solution, copernic360[1], has been developed by Kagenova[2] to provide 6-degrees-of-freedom motion in 360-degree VR by simulating translational motion (McEwen *et al.*, 2020). When using copernic360 the user is able to move within the 360-degree VR experience, e.g. by walking around or leaning forward to inspect the content in the scene, and a synthetic view is generated based on the user's position in the scene. In this manner full 6-degrees-of-freedom motion, including both translational and rotational motion, is supported. This system is designed not only to provide a more realistic and immersive experience but to also reduce the mismatch between visual and vestibular cues when in a 360-degree VR experience.

---

[1] https://kagenova.com/products/copernic360/

[2] https://kagenova.com/

To achieve this effect, copernic360 includes two main sub-systems: a back-end AI-based cloud processing system; and a front-end viewer system in the form of a game engine plugin. The AI-based cloud system processes 360-degree VR video or photo content to recover an estimate of 3D geometry representing the scene. First, AI-based depth estimation techniques are used to estimate a depth map corresponding to the 360-degree content. A residual convolutional neural network (CNN) architecture is adopted, with dilated convolutions (Yu and Koltun, 2016) in order to expand the receptive field for dense predictions, like depth estimation. The architecture is similar to that previously presented by Li and colleagues (Li *et al.*, 2017), but adapted for spherical 360-degree content. Second, the estimated 360-degree depth information is then used to fit a parametric 3D geometry representing the scene, which is stored as additional meta data associated with the original content. The geometry is fitted by minimising an error cost function. Third, the viewer system uses the meta data computed by the AI-based processing system representing the parametric 3D geometry, along with the original 360-degree video or photo content, to then render a 3D textured representation of the scene. The original 360-degree content is projected onto the 3D geometry to provide a texture for the geometry. Finally, note that for 360-degree video content, the entire process is repeated for each key frame of the video. A key frame represents a frame where the content of the video has changed considerably compared to the prior frame (e.g., due to a cut between scenes or the content of a scene changing over time). Between key frames the 3D geometry is interpolated. In this manner, the user is then able to move about in the reconstructed scene with full 6-degrees-of-freedom motion and novel synthetic viewpoints of the scene are then rendered and served to the user depending on their position in the scene.

A custom scenario was adapted for VR. The scenario consisted of a beach (https://www.atmosphaeres.com/video/445/Rocky+Headland) in which participants could move around for about 10 minutes. Natural sounds were integrated such as the sounds of waves along with an auditory cue played every 20s. The auditory cue signalled participants to provide a motion sickness scores and for the researcher to record heart rate variability (see below).

## 2.4. Experimental Design and Procedure

Verbal and written instructions were given to participants at the beginning of the experiment. Data from each participant was gathered in two experimental sessions. Participants were exposed to the same scene in both standard360-VR (3-degrees-of-freedom VR) and the copernic360-VR (6-degrees-of-freedom VR) conditions. The order of experimental conditions was counterbalanced between participants. In each session, participants were asked to wear an HTC Vive head-mounted display (HMD) and asked to walk in a square pattern with a 90 degrees turn in direction at each point (Fig. 1A). Participants were trained with the walking pattern before the VR began. An auditory cue instructed participants to look down, explore the VR scene and touch the ground. It was anticipated that this would invoke greater nausea in the conventional 3-degrees-of-freedom VR conditions due to the mismatch of visuo-vestibular cues. A webcam was used to record the walking patterns as well as verbal responses.

To assess levels of cybersickness, participants completed the *Simulator Sickness Questionnaire* (SSQ; Kennedy *et al.*, 1993) following conclusion of the VR scenario. This well-established questionnaire asks participants to score 16 symptoms on a four point scale (0=none, 3=severe). The questionnaire divides symptoms of sickness into components of Nausea (SSQ-N), Disorientation (SSQ-D) and Oculomotor (SSQ-O) clusters (Kennedy *et al.*, 1993). Disorientation (SSQ-D) includes symptoms such as dizziness, vertigo and difficulty focusing. Oculomotor (SSQ-O) includes physical symptoms such as eyestrain, headache and blurred vision. Nausea (SSQ-N) is comprised of symptoms including stomach awareness, increased salivation and nausea itself. The SSQ-N, SSQ-O, and SSQ-D scores are calculated from the weighted totals using the following conversion formulas (Kennedy *et al.*, 1993):

$$\text{SSQ-N} = [\text{ Sum obtained by adding symptom scores }] \times 9.54$$
$$\text{SSQ-O} = [\text{ Sum obtained by adding symptom scores }] \times 7.58$$
$$\text{SSQ-D} = [\text{ Sum obtained by adding symptom scores }] \times 13.92$$

Scores are then summed together to obtain a single score (Total, SSQ-T = (Nausea + Oculomotor + Disorientation) x 3.74).

During VR exposure, participants were asked to perform the *Fast Motion Sickness Scale* (FMS; Keshavarz and Hecht, 2011) and give a verbal rating of the level of nausea from 0-20 (0=no nausea; 20= frank sickness).  Responses were collected every 20 seconds when an auditory cue was played during VR, before, and immediately after VR exposure.  Participants' heart rate was also monitored throughout the VR scenario.  Heart rate (HR) has been shown to increase with greater levels of sickness in VR (Kim *et al.,* 2005; Nalivaiko *et al*., 2015). Participants wore a smart watch (Mio ALPHA 2 smart watch, Mio Technology, Taipei, Taiwan) which provided continuous readings of heart rate, with measurements recorded at the same time as the FMS ratings.  Thus, heart rate was recorded once prior to commencing the VR scenario, every 20 seconds during the scenario, and once immediately following the scenario.

Participants were also asked to complete the Motion Sickness Susceptibility Questionnaire Short-form (MSSQ) and a Gaming Experience Questionnaire to ensure that motion sickness susceptibility and previous VR experience were similar across our sample.

## 2.5. Data analysis

Supporting data are available as Supplementary Material.

The data was analysed using parametric tests.  Parametric tests are a preferred method in repeated measures randomised designs in which the interest is on the effects of an intervention/treatment vs. baseline/control (Vickers, 2005). Importantly, parametric tests have generally more statistical power than nonparametric tests, and are therefore are more likely to detect a significant effect when one truly exists.

Paired t-tests were used to analyse differences between the standard360-VR (3-degrees-of-freedom VR) and copernic360-VR (6-degrees-of-freedom VR) conditions.  T-tests were applied on the SSQ scales (SSQ-N, SSQ-D, SSQ-O) and total scores (SSQ-T).

A 2 x 3 Repeated Measures ANOVA was used to explore the changes in FMS and HR measures across the time.  Average FMS value and peak FMS value during VR exposure were entered in two distinct ANOVAs comparing differences FMS scores between VR conditions (standard360-VR vs copernic360-VR) across time (Pre-VR, During-VR and Post-VR). Similarly, average HR value and peak HR value during VR exposure were entered in two distinct ANOVAs comparing differences in HR between VR conditions (standard360-VR vs copernic360-VR) across time (Pre-VR, During-VR and Post-VR). Mauchly's test was used to investigate whether the assumption of sphericity was violated, and accordingly the degrees of freedom were corrected using Greenhouse-Geisser.

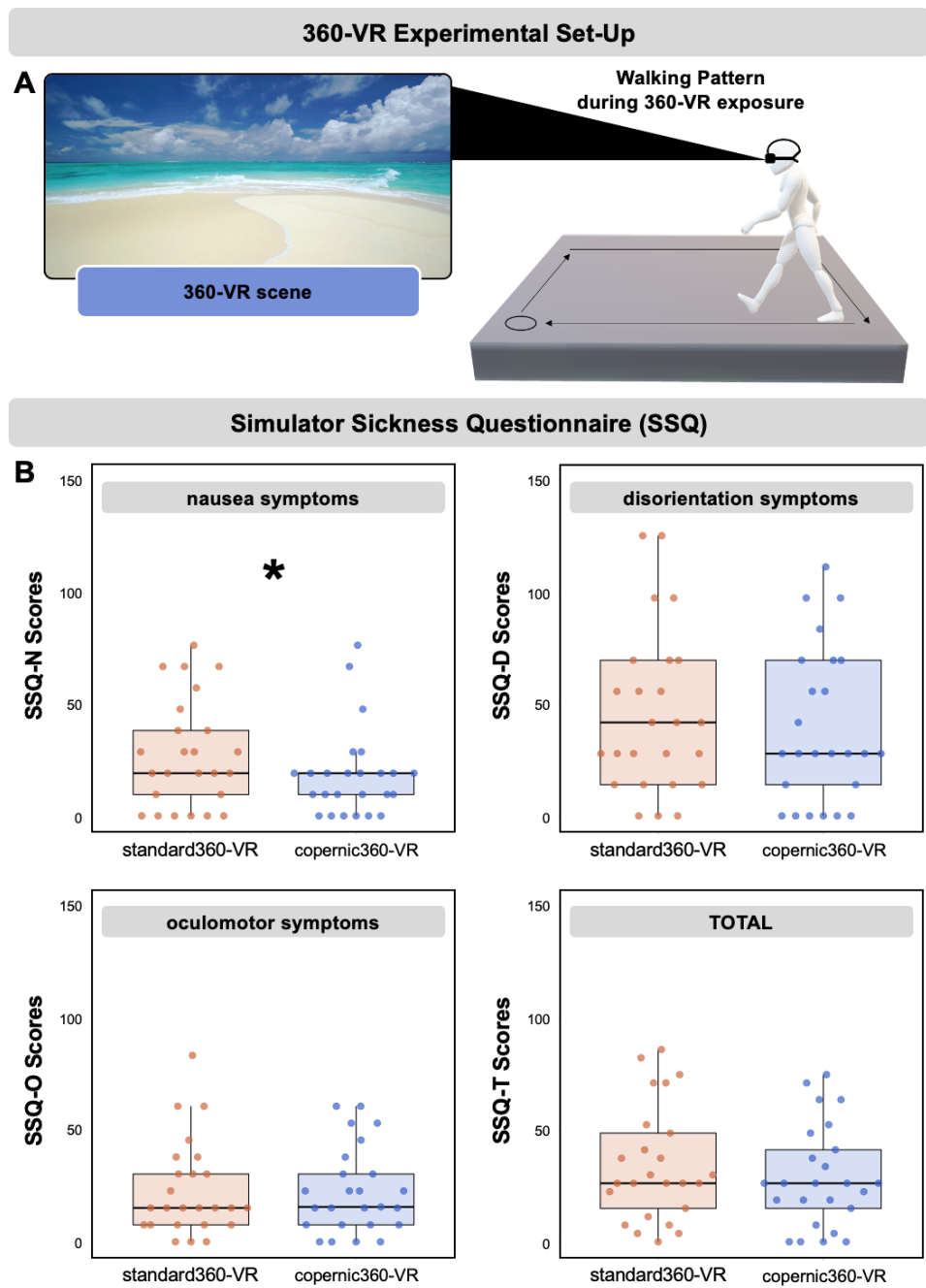

**Figure 1. Experimental Set Up and Simulator Sickness Questionnaire (SSQ) Results**

(A) Participants were exposed to a custom Virtual Reality (VR) scenario in both standard360-VR (3-degrees-of-freedom VR) and the copernic360-VR (6-degrees-of-freedom VR) setting. (B) Results showed a significant reduction in nausea on the Simulator Sickness Questionnaire (SSQ) in the copernic360-VR condition compared to standard360-VR (3-degrees-of-freedom VR). No significant differences emerged on the oculomotor and disorientation dimensions of the questionnaire. No significant difference also emerged on the total scores.

## 3. Results

SSQ scores across VR experimental conditions can be seen in Fig. 1B.

A significant reduction in nausea on the SSQ scale was observed in the copernic360-VR (6-degrees-of-freedom VR) condition compared to standard360-VR (3-degrees-of-freedom VR) ($t(24) = 2.17$, $p = .04$, Cohen's d = 0.44). Participants reported fewer symptoms of nausea in the copernic360-VR condition (M = 18.32, SD = 19.65) compared to standard360-VR (M = 27.48, SD = 24.22). There were no significant differences between VR conditions on the oculomotor ($t(24) = .128$; $p = .90$, Cohen's d = 0.02) and disorientation ($t(24) = .929$; $p = .36$, Cohen's d = 0.19) dimensions of the SSQ. No differences between 6-degrees-of-freedom VR and 3-degrees-of-freedom VR conditions emerged in the SSQ total scores ($t(24) = 1.12$, $p = .28$, Cohen's d = 0.23).

FMS scores across VR experimental conditions can be seen in Table 1. FMS scores showed no significant effect of VR condition on average FMS rating ($F(1, 24) = .32$, $p = .58$, $np2 = .013$). However, there was a significant main effect of time ($F(1.34, 32.18) = 23.64$, $p < .001$, $np2 = .50$). No interaction emerged between VR condition and time ($F(1.42, 34.06) = .52$, $p = .60$, $ηp2 = .02$). Post-hoc Bonferroni tests showed that During-VR average FMS rating (M= 3.08, SE= .62) was significantly higher than Pre-VR FMS rating (M=.20, SE= .11) ($t(24) = 5.38$, $p < .001$, Cohen's d = 1.08) across VR conditions. There was no differences between During-VR and Post-VR FMS ratings ($t(24) = -0.86$, $p = .17$). Post-VR FMS rating (M=3.34, SE= .72) was higher than Pre-VR FMS rating ($t(24) = 4.99$, $p < .001$, Cohen's d = 1). A similar analysis performed on peak FMS scores during VR revealed no significant effect of VR condition when peak FMS ratings were used ($F(1, 24) = .77$, $p = .39$, $np2 = .031$). There was a significant effect of time on FMS peak ratings ($F(1.58, 38.01) = 39.65$, $p<.001$, $np2$ .62). There was no significant interaction between VR condition and time ($F(2, 48) = .67$, $p = .52$, $ηp2 = .03$). Follow-up testing using Bonferroni correction showed that During-VR peak FMS rating (M=5.38, SE= .75) was significantly higher than Pre-VR FMS rating (M=.20, SE= .11) ($t(24) = 7.61$, $p < .001$, Cohen's d = 1.52) across VR conditions. During-VR peak FMS rating was higher than Post-VR FMS rating (M=3.34, SE= .72) ($t(24) = 4.92$, $p < .001$, Cohen's d =

0.98). Post-VR FMS rating was higher compared to Pre-VR FMS ratings (t (24) = 4.99, p < .001, Cohen's d = 1). The increase in FMS ratings indicates increased cybersickness symptoms over time. However, this increase was equally present across standard360-VR (3-degrees-of-freedom VR) and copernic360-VR (6-degrees-of-freedom VR) conditions.

HR values across VR experimental conditions can be seen in Table 2. There was no significant effect of VR condition (3-degrees-of-freedom VR vs 6-degrees-of-freedom VR) on average heart rate measures (F (1, 24) = 1.23, p = .28, $\eta p2$ = .05). A significant main effect of time (F (1.38, 33.19) = 4.22, p =.036, $\eta p2$= .15) emerged. There was no significant interaction between VR condition and time (F (2, 48) = 1.13, p = .33, $\eta p2$ = .05). Post-hoc Bonferroni tests revealed that Post-VR HR measures (M=102.28, SE= 2.95) were higher than During-VR average HR measures (t (24) = 2.54, p = .02, Cohen's d = 0.51). No differences emerged between During-VR average HR measures (M= 98.33, SE= 2.28) and Pre-VR HR measures (M=96.14, SE= 2.83) (t (24) = 1.12, p = .27), and between Post-VR HR measures (M=102.28, SE= 2.95) and Pre-VR heart rate measures (t (24) = 2.24, p = .04). A similar analysis performed on peak HR values during VR revealed no significant effect of VR condition when peak HR values were used (F (1, 24) = 1.66, p = .21, $np2$ = .065). There was a significant effect of time on HR peak values (F (2, 48) = 19.74, p<.001, $np2$ .45). There was no significant interaction between VR condition and time (F (2, 48) = .37, p = .70, $\eta p2$ = .02). Follow-up testing using Bonferroni correction showed that During-VR the peak HR measures (M= 111.3, SE= 2.45) was significantly higher compared to both Pre-VR HR measures (M=96.14, SE= 2.83) (t (24) = 7.24, p < .001, Cohen's d = 1.45) and Post-VR HR measures (M=102.28, SE= 2.95) (t (24) = 3.75, p = .001, Cohen's d = 0.75). Post-VR HR measures were no different from Pre-VR HR measures (t (24) = 2.24, p = .04). Although heart rate increased during VR exposure, no differences emerged between standard360-VR (3-degrees-of-freedom VR) and copernic360-VR (6-degrees-of-freedom VR) conditions.

**Table 1. Mean (SE) for FMS scores across time in both experimental conditions.**

|  | standard360-VR | copernic360-VR |
|---|---|---|
| Pre-VR | 0.16 (0.09) | 0.24 (0.20) |
| During-VR (average measure) | 3.20 (0.67) | 2.97 (0.67) |
| During-VR (peak measure) | 5.80 (0.93) | 4.96 (0.83) |
| Post-VR | 3.60 (0.86) | 3.08 (0.74) |

**Table 2. Mean (SE) for HR values across time in both experimental conditions.**

|  | standard360-VR | copernic360-VR |
|---|---|---|
| Pre-VR | 95 (3.19) | 97.28 (3.12) |
| During-VR (average measure) | 98.22 (2.25) | 98.45 (2.56) |
| During-VR (peak measure) | 110.52 (2.14) | 112.08 (3.11) |
| Post-VR | 100.44 (3.35) | 104.12 (3.01) |

## 4. Discussion

Despite the improvements in VR hardware, cybersickness is a predominant factor affecting VR experience and feeling of immersion. The mismatch between visuo-vestibular cues can induce compelling feelings of nausea and discomfort. In this study, we focused on a particular VR experience in which videos are captured using omnidirectional cameras which enable the filming of an entire 360 degrees scene. In 360-degree VR the user is able to look and move around the entire scene. However, 360-degree VR only supports 3-degrees-of-freedom rotational motion, which might easily conflict with the level of precise information transmitted by the vestibular organs as soon as the user moves their head. That is, traditional 360-degree VR is a unique setting offering minimal dynamic visual inputs combined with a flow of translational vestibular inputs. Here we investigated whether cybersickness in 360-degree VR can be reduced by a novel AI-based software (copernic360). Both explicit (simulator sickness questionnaire and fast motion sickness rating) and implicit (heart rate) measures were used to quantify the changes in cybersickness. Our results showed a significant reduction in nausea symptoms on the Simulator Sickness Questionnaire (Kennedy *et al.,* 1993) when using copernic360-VR compared to traditional 360-degree VR. Importantly, the oculomotor and disorientation dimensions were not significantly different between VR conditions. However, the significant reduction in nausea symptoms demonstrates the potential of the copernic360-VR software in improving the VR experience. As anticipated, we observed an increase in motion sickness ratings and heart rate towards the end of VR exposure. While there were no significant differences between the copernic360 VR and traditional 360-degree VR conditions for these measures, the increase appears more marginal in copernic360-VR.

Cybersickness has been often assessed via subjective self-reports. The Simulator Sickness Questionnaire (Kennedy *et al.,* 1993) is probably the most frequently used. This questionnaire breaks down motion sickness symptoms into three main categories: disorientation (including dizziness, vertigo and difficulty focusing), oculomotor (eyestrain, headache, and blurred vision) and nausea (stomach awareness, increased salivation, and nausea itself). In typical VR scenarios in which the user is experiencing vection while vestibular cues are signaling zero angular or translational acceleration, cybersickness is characterised by a D>N>O profile in which the most severe and frequent symptoms are observed in the disorientation domain, followed by nausea symptoms, and least oculomotor symptoms (Rebenitsch and Owen, 2016; Stanney *et al.*, 2003). This pattern of symptoms has been explained by a sensory mismatch between visual and vestibular cues in VR which may lead to difficulties in forming and updating spatial maps of the external virtual environment and the organism (Lackner and DiZio, 2005), leaving the VR user disoriented. Similar to previous studies, here we observed an overall D>N>O profile across experimental conditions (standard360-VR and copernic360-VR). Although copernic360-VR numerically reduced sickness scores of disorientation, nausea and oculomotor symptoms, it only significantly improves the nauseogenic aspect of the 360-degree VR experience. Importantly, nausea is consistently experienced in 360-degree settings (Gavgani *et al.*, 2017) and it has been associated with significant user drop out (Balk *et al.*, 2013; Ehrlich and Kolasinski, 1998). The congruency between visual and vestibular signals induced by the copernic360-VR technology allowed for a reduction in self-reported nausea in the current study. Reducing this key symptom allows for a more immersive and interactive virtual experience. This new technology may offer an alternative to physical setups, artificial stimulation and sensory habituation protocols (Ng *et al.*, 2020).

Importantly, the scores obtained in the Simulator Sickness Questionnaire (Kennedy *et al.,* 1993) for the standard 360-degree VR condition are comparable to previous studies. Nausea levels have been reported ranging from 14.91 to 30.21, oculomotor symptoms varied from 15.30 to 25.74, and disorientation was reported markedly higher from 21.75 to 41.47 (Ehrlich *et al.*, 1998; Kolasinski and Gilson, 1998; Stanney and Kennedy, 1998). The current VR experience was characterised by a similar level of nauseogenicity. However, several factors influence cybersickness symptoms, and direct comparison between studies might not be straightforward. Not surprisingly, people susceptible to motion sickness are more likely to experience cybersickness (Rebenitsch and Owen, 2014). Symptom severity and incidence increases with age (Arns *et al.*, 2005). Some studies have indicated that females are more susceptible to cybersickness than males (Curry *et al.*, 2020; Kim *et al.*, 2020). Further, the technical characteristics of the VR equipment such as the weight of the headset as well as the VR environment might play a key role.

We measured cybersickness self-reports and heart rate variability during VR exposure. The fast motion sickness scale provides continuous data and measures of sickness during the presentation of a sickness-inducing situation, not necessary cybersickness (Keshavarz and Hecht, 2011). As expected, we observed an increase in cybersickness scores over the ten minutes of VR exposure. However, the scores were not significantly influenced by the copernic360-VR technology. It is important to note that fast motion sickness scores were overall very low and several participants never rated above zero. Our VR scenario consisted of a realistic beach scene accompanied with the soft sound of waves. While a multimodal exposure to visual and audio cues in VR may have increased a sense of presence typically leading to more sickness discomfort (Cooper *et al.*, 2018; Kim *et al.,* 2014), the calming content of our scenario may have masked feelings of cybersickness (Amores *et al.*, 2018). Accordingly, emotional and physical discomfort have been related to the Simulator Sickness Questionnaire measures (Somrak *et al.*, 2019). This might suggest that the VR scenario developed for this experiment was not particularly sickness-inducing, or more in general 360 VR is not as challenging as traditional VR. That is, the content of the VR may have reduced or counteracted against feelings of cybersickness. Further, the VR exposure is markedly less disorientating when compared to rollercoaster or first-person shooter scenarios. This may also explain the lack of significance found on the SSQ-D. Lastly, the duration of VR exposure could be extended. Increased duration has been associated with cybersickness (Moss and Muth, 2011) with 10-20 mins of exposure leading to cybersickness (Häkkinen *et al.,* 2019). Future studies should consider using a more conflicting VR scenario involving longer duration and dynamic movements. For instance, one could use a motion platform to dynamically change the tilt of the floor paired with a 360-degree video from aboard a ship. A longer duration with greater mismatch might allow the benefits of 6-degrees-of-freedom VR to be tested more thoroughly.

While some studies have reported an increased heart rate in VR (Nalivaiko *et al.*, 2015), others report conflicting findings. For instance, when participants were exposed to VR across three days, there were no significant differences found in heart rate measures (Gavgani *et al*., 2017). This might suggest a form of habituation when participants are repeatedly exposed to VR over a period of time, which may have happened in our experiment. An effect of coperic-360-VR on heart rate measure of sickness during VR could have been potentially masked by the content of the VR scenario (Guna *et al*., 2020). We cannot exclude that participants may have also adapted to the walking pattern, perhaps influencing the saliency of vestibular cues during the VR which might have in turn reduced sensory conflict and cybersickness. Along with diversifying the walking pattern to reduce the effects of habituation, more interaction with the virtual environment could be encouraged. Increased interaction through grasping movement or changes in head movement may induce greater mismatch between visual and vestibular cues leading to cybersickness. Thus, future studies should consider integrating head movements to increase the disorientation effects of VR and the benefits of 6-degrees-of-freedom VR compared to 3-degrees-of-freedom VR.

When considering the benefits of copernic360-VR, only self-reported nausea was found to differ statistically. Nausea symptoms are one of the most challenging aspects of cybersickness. Importantly, our results showed that copernic360 significantly reduced the feeling of nausea. No differences were found in the other sickness domains as well as in physiological measures. Using a more conflicting VR scenario for an extended period of time is recommended. As far as we are aware, no other approaches have implemented an AI-based 6-degrees-of-freedom technology to reduce cybersickness. Here, we offer new insights into its benefits and encourage future studies to explore the components of 3-degrees-of-freedom VR vs 6-degrees-of-freedom VR experiences.

As VR continues to improve, more needs to be done to combat the unwanted effects of cybersickness. This study demonstrated the potential of a novel AI-based software to increase the congruency between visual and vestibular cues. Crucially, feelings of self-reported nausea were reduced implying weakened feelings of cybersickness. Reducing the nausea and discomfort perceived by individuals is crucial in allowing the application and further utility of VR in educational, medicinal, cultural and entertainment settings.

**Acknowledgements**

This work was supported by a UKRI StoryFutures R&D on Demand awarded to ERF.

**Competing Interests**

The authors declared that they had no conflicts of interest with respect to their authorship or the publication of this article.

**Author Contribution**

I.A. and E.F. designed the experiment and wrote the manuscript. P.D.M, M.E. and J.D.M designed the VR technology and implemented the VR scenario. I.A. collected the data. J.D.M. reviewed the manuscript, giving a critical revision of it. All Authors gave the final approval of the version to be published.

**Figure Caption**

**Figure 1. Experimental Set Up and Simulator Sickness Questionnaire (SSQ) Results**

(A) Participants were exposed to a custom Virtual Reality (VR) scenario in both standard360-VR (3-degrees-of-freedom VR) and the copernic360-VR (6-degrees-of-freedom VR) setting. (B) Results showed a significant reduction in nausea on the Simulator Sickness Questionnaire (SSQ) in the copernic360-VR condition compared to standard360-VR (3-degrees-of-freedom VR). No significant differences emerged on the oculomotor and disorientation dimensions of the questionnaire. No significant difference also emerged on the total scores.